\documentclass[11pt]{article}
\usepackage{lineno}
\usepackage{graphicx} 
\usepackage{amsmath,amssymb} 
\usepackage{natbib}
\usepackage{authblk}

\begin{document}

\title{Epistatic strength, modularity, and locus heterogeneity shape the number of local optima in fitness landscapes}

\author[1,2]{Mahan Ghafari}
\author[3]{Alejandro Castro Cabrera}
\author[3]{Alejandro Lage-Castellanos}
\author[4]{Guillaume Achaz}
\author[5]{Joachim Krug}
\author[1]{Luca Ferretti}

\affil[1]{Pandemic Sciences Institute and Big Data Institute, Nuffield Department of Medicine, University of Oxford, United Kingdom}
\affil[2]{Department of Biology, University of Oxford, Oxford, United Kingdom}
\affil[3]{Group of Complex Systems and Statistical Mechanics, Physics Faculty, University of Havana, Cuba}
\affil[4]{Stochastic Models for the Inference of Life Evolution (SMILE), CIRB, Collège de France, Université PSL, CNRS, INSERM, 75005 Paris, France \& Université Paris-Cité, Paris, France}
\affil[5]{Institute for Biological Physics, University of Cologne, Köln, Germany}
\affil[*]{Corresponding author: Luca Ferretti, luca.ferretti@gmail.com, luca.ferretti@ndm.ox.ac.uk}

\date{}
\maketitle

\begin{abstract}
Fitness landscapes provide a quantitative framework for understanding how natural selection shapes evolutionary trajectories. A central feature of these landscapes is their number of local optima, which determines whether fitness-increasing evolution can proceed towards a global optimum or become trapped on suboptimal peaks. Although multiple peaks are known to require reciprocal sign epistasis, the quantitative relationship between epistasis and number of peaks remains incompletely understood. Here, we show that for a broad class of unstructured fitness landscapes, i.e. isotropic Gaussian random fields, the expected number of local optima is determined by a single local measure of epistasis: the correlation of fitness effects. This provides a baseline prediction for the number of peaks in typical unstructured landscapes and links peak density directly to the amount of reciprocal sign epistasis. This baseline changes when epistatic interactions are structured. We show that clustering interactions 
within blocks of loci slightly increases the number of local optima. 
In contrast, strong heterogeneity between loci, where only a small subset of loci participate in epistatic interactions, causes the number of peaks to collapse. These results show that the number of local optima is governed not only by the overall strength of epistasis, but also by how epistatic interactions are distributed across the genotype space. Our framework therefore reconciles the central role of reciprocal sign epistasis with the observation that landscapes with similar amounts of epistasis can differ substantially in ruggedness, and provides a guide to the range of peak numbers expected in typical landscapes.
\end{abstract}

\noindent\textbf{Keywords:} fitness landscapes; local optima; epistasis; reciprocal sign epistasis; Gaussian random fields; NK model; ruggedness


\section{Introduction}

Fitness landscapes are a central concept in evolutionary genetics, describing the mapping between genotypes and fitness \citep{Wright1932,svensson2012adaptive,srivastava2026}. They provide a natural framework for asking how populations move through genotype space under evolutionary forces and why evolution may sometimes be predictable and at other times strongly constrained \citep{de2014empirical}. A key topological feature of a fitness landscape is its number of local optima. Local optima, or fitness peaks, are genotypes that have higher fitness than their mutational neighbors. Their number is often used as a measure of the ``ruggedness'' or complexity of a landscape because it determines whether fitness-increasing trajectories can reach a unique global optimum or instead become trapped on suboptimal peaks. 

Epistasis, i.e. interactions between mutations that cause their fitness effects to depend on the genetic background, is a key feature of fitness landscapes \citep{Domingo2019,bank2022epistasis,diaz2023global}. It is well understood that the number of local peaks is determined by the amount and structure of epistatic interactions. There is a large body of literature focused on the expected number of local fitness peaks in many epistatic models such as House of Cards/Random Energy Models \citep{Kauffman1987,macken1989protein}, NK models \citep{Kauffman1989,Weinberger1991,perelson1995protein,hwang2018universality}, and Rough Mount Fuji models \citep{Aita2000,neidhart2014adaptation}, see \cite{pahujani2025complexity} for a recent review.

A particularly important form of local epistasis is sign epistasis, where the effect of a mutation changes from beneficial to deleterious, or vice versa, depending on the presence of another mutation \citep{weinreich2005perspective}. If both mutations in a pair change sign in each other's presence, the interaction is termed reciprocal sign epistasis \citep{Poelwijk2007}. Reciprocal sign epistasis has a special relationship with local optima: it is a necessary condition for the existence of multiple peaks \citep{Poelwijk2011}. More precisely, a landscape with more than one peak must contain reciprocal sign epistasis, and a landscape with $K$ peaks must contain at least $K-1$ reciprocal sign epistatic motifs \citep{riehl2022occurrences,saona2022relation}. Yet this condition is not sufficient: the presence of reciprocal sign epistasis alone does not determine how many peaks a landscape will have \citep{Crona2013}. Recent work from our group has shown that simple sign epistasis, in which only one of two mutations changes sign, affects evolutionary trajectories in a different way: it is associated with longer indirect paths and evolutionary detours but not necessarily with multiple peaks \citep{Ribeca2026}. This motivates a more quantitative question: beyond the existence of reciprocal sign epistasis, what determines the typical number of peaks in a landscape?

The case of NK models is a very interesting example. NK models allow for different interaction structures for the same amount and strength of epistatic interactions. It had been conjectured that the number of local peaks was universal with respect to the choice of interactions \citep{Weinberger1991}, but it was subsequently shown that biallelic landscapes with different choices of interactions (such as random and mean-field interactions versus block and adjacent interactions) have very different numbers of local peaks \citep{hwang2018universality}. However, this observation is restricted to a particular class of landscape models. A more general open question is: what range of numbers, or densities, of local peaks should we expect in \emph{typical} fitness landscapes? How does this range depend on epistasis or reciprocal sign epistasis, and what other properties determine it?

Here we develop a unified framework for understanding how the number of local optima depends on the strength and structure of epistatic interactions. We first consider a broad class of unstructured landscapes, often referred to as isotropic Gaussian random fields. These models are unstructured in the sense that all loci and alleles are exchangeable and epistatic interactions are not concentrated among particular subsets of loci. For this class, we show that the expected number of local optima depends only on the number of loci and a single epistasis parameter, $\gamma$, defined as the correlation between the fitness effects of the same mutation on neighbouring genetic backgrounds \citep{Ferretti2016}. Equivalently, this relationship can be expressed in terms of the expected amount of reciprocal sign epistasis.

We then extend this baseline model to landscapes with structured epistasis. We focus on two forms of structure that preserve the average amount of epistasis but distribute it differently across loci: clustering and heterogeneity. By clustering (or modularity), we mean that epistatic interactions are concentrated within groups of loci, so that loci interact preferentially with other loci in the same block. By locus heterogeneity, we mean that loci differ in whether they participate in epistasis at all, with some loci contributing mainly additive effects while others carry most of the epistatic interactions. Thus, clustering describes the organisation of interactions among epistatic loci, whereas heterogeneity describes unequal participation of loci in epistasis.

First, we consider clustering of interactions, where epistatic interactions are concentrated within blocks of loci. This structure tends to increase the number of local optima relative to unstructured landscapes. 
Second, we consider heterogeneity between loci, where only a subset of loci participate in epistatic interactions while the remaining loci are effectively additive. Moderate heterogeneity can have mixed effects because it both reduces the effective dimensionality of the rugged component and increases interaction strength within that component. However, strong heterogeneity has the opposite effect to clustering and leads to a collapse in the number of local optima even when the average amount of epistasis is held fixed.

Together, these results show that the typical number of local optima is mostly determined by two ingredients, namely the strength of local epistasis and the heterogeneity of interaction participation across loci, with a minor role for the clustering of interactions. The unstructured Gaussian model provides a natural baseline, while clustered and heterogeneous models define upper and lower deviations from this baseline. This framework explains why reciprocal sign epistasis is necessary for multiple peaks but insufficient to predict their number, and why landscapes with similar epistasis can differ substantially in ruggedness.


\section{Materials and Methods}
\label{sec:materials:methods}

\subsection{Fitness landscapes and local optima}

We consider fitness landscapes defined on a genotype space of sequences of length $L$, with $A$ possible alleles at each locus. A genotype is denoted by $g$, and its fitness by $f(g)$. A single mutational neighbour of $g$ is a genotype differing from $g$ at one locus. A genotype $g$ is a local optimum, or local fitness peak, if its fitness is greater than that of all its neighbours one mutation away.

For a mutation to allele $a$ at locus $i$ from genotype $g$, we denote the fitness effect by
\begin{equation*}
\Delta_{i,a} f(g) = f(g_{[i,a]}) - f(g),
\end{equation*}
or more simply, for biallelic landscapes (and with a slight abuse of notation, even for multiallelic ones),
\begin{equation*}
\Delta_i f(g) = f(g_{[i]}) - f(g),
\end{equation*}
where $g_{[i]}$ is the genotype obtained by mutating locus $i$ to allele $a$. Epistasis occurs when the effect of a mutation depends on the genetic background. In local terms, two mutations at loci $i$ and $j$ interact epistatically if
\begin{equation*}
\Delta_{i,a} f(g) \neq \Delta_{i,a} f(g_{[j,a']}).
\end{equation*}
We quantify local epistasis using the correlation of fitness effects \citep{Ferretti2016},
\begin{equation}
\gamma = \mathrm{Cor}[\Delta_i f(g),\, \Delta_i f(g_j)],
\end{equation}
where $g$ and $g_j$ differ by a single mutation at a locus distinct from $i$. In the absence of epistasis, $\gamma = 1$. As epistasis increases, $\gamma$ decreases. This parameter captures how stable the fitness effect of a mutation is across nearby genetic backgrounds and is therefore directly linked to the local geometry of the landscape.

\subsection{Sign epistasis and reciprocal sign epistasis}

For pairs of mutations, we distinguish magnitude epistasis (ME), simple sign epistasis (SSE), and reciprocal sign epistasis (RSE) \citep{weinreich2005perspective,Poelwijk2007}. ME occurs when the magnitude, but not the sign, of a mutation's effect changes across genetic backgrounds. SSE occurs when one mutation changes sign depending on the presence of the other. RSE occurs when both mutations change sign depending on the presence or absence of the other.

We denote the fraction of RSE motifs by $\phi_{\mathrm{rs}}$. RSE is directly related to the existence of multiple peaks, but $\phi_{\mathrm{rs}}$ alone does not completely determine the number of peaks, because the spatial organisation of RSE motifs across the genotype space also matters. In unstructured Gaussian landscapes, however, $\phi_{\mathrm{rs}}$ can be expressed as a function of $\gamma$ \citep{Ribeca2026}, making it possible to relate peak density, local epistasis, and RSE within a single analytical framework.

\subsection{Unstructured landscape models}
In this section, we introduce a large and important class of typical unstructured landscapes, i.e.\ isotropic Gaussian random field landscape models.

Epistatic interactions can involve different numbers and choices of loci with different strength. This variety and complexity makes it very difficult to obtain general results for large classes of models. We propose two natural simplifications: (i) We consider isotropic landscapes where all genotypes and loci behave in a similar way. Interactions do not occur preferentially among specific sets of loci, but are distributed randomly across all sets of loci. In this sense, interactions are unstructured. (ii) If the interactions are the result of a combination of a large number of small effects, their strength tends to be normally distributed. The combination of multiple such interactions gives rise to a correlated Gaussian landscape called an isotropic Gaussian random field landscape. 

These landscapes are defined by assigning correlated Gaussian fitness values to genotypes in a way that is invariant under permutations of loci and alleles. Equivalently, 
the fitness function can be represented through a Fourier expansion over the genotype network \citep{weinberger1991fourier}, with epistatic coefficients drawn from independent Gaussian distributions whose variances $V_k$ determine the epistatic spectrum \citep{stadler1999random,Neidhart2013}. These Gaussian distributions are centered for all coefficients apart from the constant one, and their variances depend only on the order of the interaction (i.e. the number of loci involved). Therefore, up to an additive constant on fitness, such landscape models are defined by $L-1$ parameters corresponding to the ``power spectrum'' of the Fourier expansion.

These models are ``unstructured'' because no particular loci or groups of loci are privileged: epistatic interactions are distributed homogeneously across the genotype space. Despite this symmetry, the epistatic spectrum can be arbitrarily complex, including pairwise and higher-order interactions of different strengths. This wide class of landscapes includes several models of interest. Unstructured (mean-field) NK models belong to this class and represent good approximations of NK models with random neighbours \citep{hwang2018universality}. Rough Mount Fuji models with a normally distributed additive component also fall into this class, as well as models inspired by spin glasses and statistical physics such as the Sherrington-Kirkpatrick model \citep{stein1992}.

For this class of models, there is a simple relation between the expected fraction of reciprocal sign epistasis and the expected correlation of fitness effects $\gamma$ \citep{Ribeca2026}:
\begin{equation}
\phi_{rs}=\frac{2}{\pi}\arcsin\left(\frac{1-\gamma}{2}\right) \label{eq:rs}.   
\end{equation}

There is also a simple relation between $\gamma$ and the power spectrum $V_k$ of the model, i.e. the variance of the Gaussian-distributed Fourier coefficients of order $k$. This relation can be obtained using the known relation between the fitness autocorrelation functions at distance $d$, $\rho_d$ and the variances of the coefficients of order $k$, $V_k$ through Krawtchouk polynomials \citep{Stadler1996,Neidhart2013}, and the alternative definition $\gamma=\frac{\rho_1-\rho_2}{1-\rho_1}$ \citep{Ferretti2016}. For example, for the biallelic case, the relation is
\begin{equation}
\gamma=1-2\frac{\sum_{k=1}^{L}k(k-1){L\choose k} V_k}{\sum_{k=1}^{L}k(L-1){L\choose k} V_k}\label{eq:gamma_spectrum}.
\end{equation}

The key result of the present study is that, for this broad class of landscapes, the expected number of local optima depends on the epistatic spectrum only through the local epistasis parameter $\gamma$. Equivalently, given the above equation (\ref{eq:rs}), it can be expressed as a universal function of $\phi_{rs}$.

Note that landscapes with $\gamma<0$ are rare in practice. The reason is that these landscapes have a complex interaction structure dominated by purely higher-order interactions, i.e. compensatory interactions of order $k>(L+1)/2$, as it can be seen immediately from equation (\ref{eq:gamma_spectrum}). These interactions also imply that the occurrence of a mutation tends to revert the fitness effect of many other mutations in the landscape. There is no known biological mechanism that generates these interactions on a regular basis. Therefore, while all our results for pure unstructured Gaussian landscapes are generally valid for any isotropic Gaussian random field model, we consider these landscapes to be typical only for $\gamma\geq 0$. 

Note that for these typical unstructured Gaussian landscapes, the relation between $\phi_{rs}$ and $\gamma$ is basically linear up to a 5\% error, since for $\gamma\geq 0$ 
\begin{equation}
\frac{1-\gamma}{\pi} \leq \phi_{rs} \leq \frac{1-\gamma}{3}\simeq \frac{1-\gamma}{\pi}\times 1.047.
\end{equation}



\subsection{Models with heterogeneity and clustering of epistatic interactions}

In this section, we build a more general class of models that include heterogeneity between loci and clustering of epistatic interactions.

\subsubsection{Models with clustered interactions:}
The most extreme structure on the interactions that can be imposed while preserving the condition that no locus or allele should have any special property is a block structure. Namely, loci are assigned to $B$ blocks of approximately identical size $\sim L/B$. There are no interactions between mutations in different blocks, while all loci within each block have unstructured interactions characterised by an arbitrary spectrum with epistasis $\gamma_b\geq 0$. We characterise the clustering using the parameter $\beta\in [0,1]$ denoting the number of excess blocks relative to its maximum:
\begin{equation}
\beta=\frac{B-1}{L-1}\ \ \Rightarrow\ \ B=1+\beta(L-1)
\end{equation}
For large landscapes, the inverse of this parameter $1/\beta$ is effectively the number of loci in each block.

\subsubsection{Models with heterogeneity between loci:} If we drop the requirements that all loci interact the same way, the most extreme asymmetry is to include a subset of non-interacting loci. The fraction of loci interacting epistatically is denoted by $f_e$. Therefore, these landscapes represent a combination of $L(1-f_e)$ non-interacting loci and $Lf_e$ loci with unstructured interactions characterised by an arbitrary spectrum with epistasis $\gamma_b\geq 0$. We assume that the mean squared fitness effect $E[(\Delta f)^2]$ is the same for both interacting and non-interacting loci. We also define a heterogeneity parameter $h\in [0,\infty]$ as the ratio between the number of non-interacting and interacting loci, i.e.
\begin{equation}
h=\frac{1}{f_e}-1\ \ \Rightarrow\ \ f_e=\frac{1}{1+h}
\end{equation}

\subsubsection{Full models:}
We can combine the two types of heterogeneity in a general model. This model has a fraction $1-f_e=\frac{h}{1+h}$ of loci that do not interact at all, while the other $f_e=\frac{1}{1+h}$ loci have a block structure with $B$ blocks of loci, and no epistatic interactions between different blocks. Within each block, interactions follow an unstructured Gaussian model with arbitrary spectrum with epistasis $\gamma_b\geq 0$. Given a genotype $g=(a_1\ldots a_{1-Lf_e},a_{1,1}\ldots a_{B,Lf_e/B})$:
\begin{equation}
f(g)=\sum_{i=1}^{L(1-f_e)}f_{\text{linear}}(a_i)+\sum_{J=1}^Bf_{\text{unstructured}}(a_{J,1}\ldots a_{J,Lf_e/B})
\end{equation}
where again we assume that $E[(f_{\text{linear}}(a)-f_{\text{linear}}(a'))^2]=E[(\Delta_{} f_{\text{unstructured}}^2]$.

The overall measure of local epistasis is 
\begin{equation}
1-\gamma=\frac{f_e(f_e/B-1/L)}{1-1/L}(1-\gamma_b)    
\end{equation}
and the condition $\gamma_b\geq 0$ for typical landscapes means that typical values for the parameters $\gamma,B,f_e$ should satisfy the constraint
\begin{equation}
(1-\gamma)\frac{1-1/L}{f_e(f_e/B-1/L)}\leq 1    
\end{equation}
or, if we redefine the clustering parameter $\beta$ for this model as
\begin{equation}
\beta=\frac{B-1}{Lf_e-1}\ \ \Rightarrow\ \ B=1+\beta(Lf_e-1)
\end{equation}
then should satisfy the equivalent constraint in terms of $\gamma,\beta,h$:
\begin{equation}
(1-\gamma)\frac{(1+h)(1-1/L)}{(1/(1+\beta(L/(1+h)-1))-1/L)}\leq 1    
\end{equation}
Also, in this model, the fraction of RSE is determined by $\gamma$ through the relation:
\begin{equation}
\phi_{rs}=\frac{2f_e(f_e/B-1/L)}{\pi(1-1/L)}\arcsin\left(\frac{(1-1/L)(1-\gamma)}{f_e(f_e/B-1/L)}\right)\simeq \frac{(1-\gamma)}{3}
\end{equation}

\noindent which is approximately linear in the typical weak-epistasis regime.

\subsection{Density of peaks for long sequences}
For sequences of finite length, we report the number of local fitness peaks $N_{peaks}$ as a function of $L,A,\gamma,\beta,h$. The density of peaks can simply be computed by $\pi_{peaks}=N_{peaks}/A^L$.

However, for long sequences $L\rightarrow\infty$, we consider the realistic case where the effective number of strong interactions per site $k=L(1-\gamma)$ is fixed \citep{hwang2018universality}. In this limit, the scaling for the density of peaks is often $\pi_{peaks}\sim e^{\lambda L}$ with $\lambda<0$, therefore we will report the value of the exponential rate $\lambda=\lim_{L\rightarrow\infty}\log(\pi_{peaks)})/L$.

The expected number of local optima then scales as $N_{peaks}\sim A^L e^{\lambda L}=e^{(\log A + \lambda)L}$.

\subsection{Analysis of empirical landscapes}

We compare the theoretical predictions for unstructured landscapes with a collection of experimentally characterised biallelic fitness landscapes and their complete sublandscapes \citep{} already presented in \citep{,}. For each empirical landscape, we compute the number of local optima, measures of global roughness (such as the roughness/slope ratio and the fraction of fitness variance due to nonlinear components in the Fourier spectrum \citep{Szendro2013}), local epistasis ($\gamma$ and the equivalent measure for fitness graphs $\gamma^*$ \citep{Ferretti2016}), and sign epistasis including the fraction of reciprocal sign epistasis motifs. These comparisons allow us to test whether empirical landscapes are better explained by local epistasis than by global roughness, and whether their numbers of peaks is close to the expectation predicted by unstructured Gaussian models.

\section{Results}

\subsection{The number of local optima depends on local epistasis}

In this section, we discuss the general evidence for the dependence of the number of local optima on epistasis, clarifying the relationship between different notions of epistasis and the number of peaks. 

Global measures of epistasis, or ``roughness'', measure the overall deviation from additivity. However, the number of local optima is determined by local comparisons between neighbouring genotypes. In the absence of fitness ties, a genotype is a peak only if all one-step mutations away from it are deleterious. Therefore, peak number depends directly on the distribution and correlation of local fitness effects, rather than roughness. This distinction explains why landscapes with similar global roughness can have very different numbers of peaks. If nonlinear variation is distributed in a way that changes the local signs of mutational effects, the landscape can become highly rugged. If the same amount of nonlinear variation preserves the relative signs of local effects, the landscape can remain comparatively smooth.

\begin{figure}
\centering
\includegraphics[width=\linewidth]{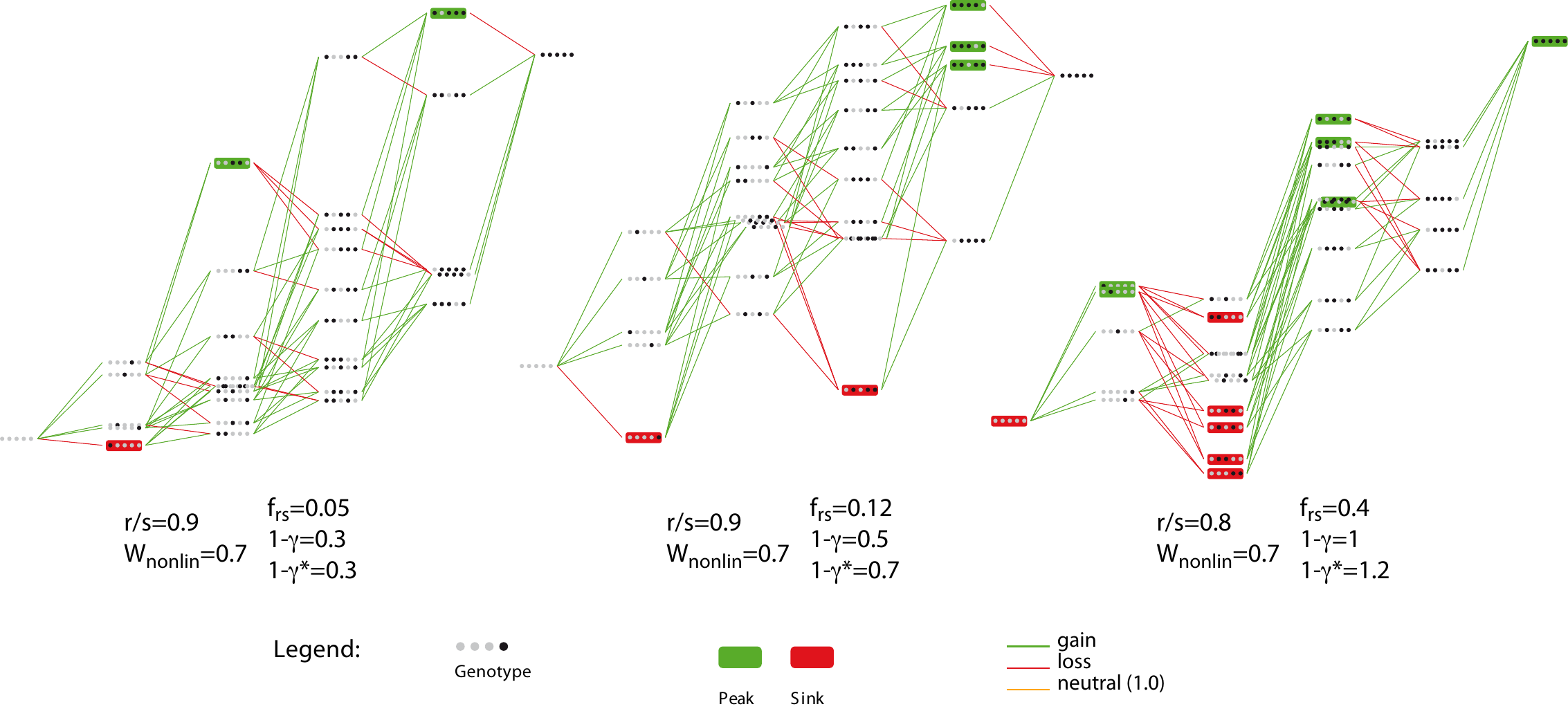}
\caption{An illustration of the argument that local epistasis, not roughness, determines the number of peaks. These three landscapes have approximately the same amount of roughness, but their nonlinear spectrum is very different (pairwise interactions for the first, random House-of-Cards interactions for the second and maximally epistatic interactions for the third), hence their number of local maxima and minima is completely different and strongly correlated with measures of local epistasis.}
\label{fig_model_landscapes}
\end{figure}

This is illustrated in Figure~\ref{fig_model_landscapes}, which shows three landscapes with nearly identical roughness but different epistatic spectra and, correspondingly, very different numbers of local optima.

\begin{figure}
\centering
\includegraphics[width=\linewidth]{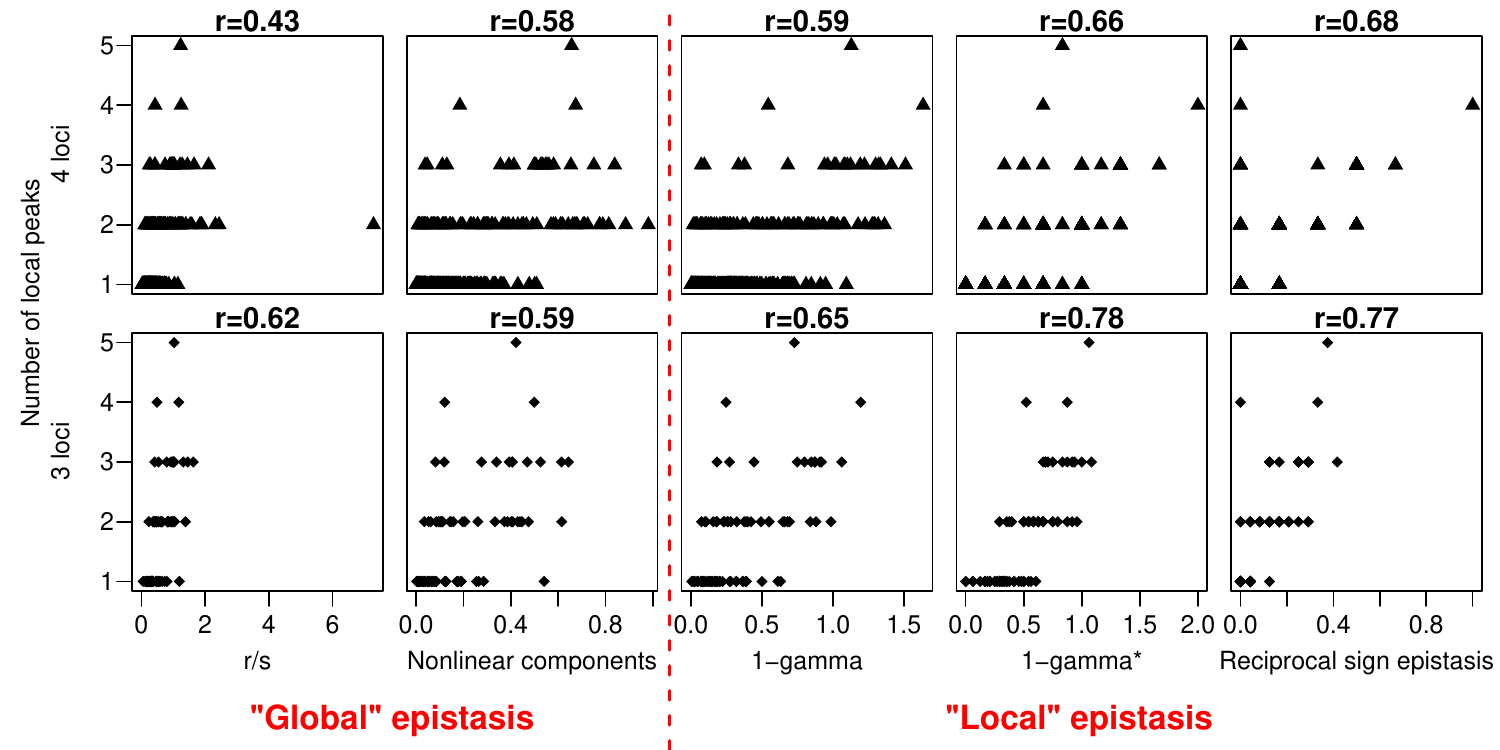}
\caption{Correlation between the number of peaks/sinks and several measures of local and global roughness across all sublandscapes of size $L=3,4$ of some empirical biallelic landscapes.}
\label{fig_empirical}
\end{figure}

This is also supported by the analysis of experimentally characterised biallelic (sub)landscapes of size $L=3$ and $4$. The number of optima across these empirical landscapes is much more correlated with local epistasis (Pearson's $r^2=0.35-0.61$) than with global roughness (Pearson's $r^2=0.18-0.38$), as shown in Figure~\ref{fig_empirical}.

The relationship between RSE and local optima provides a second very strong argument for focusing on local epistasis. Multiple peaks require RSE \citep{Poelwijk2011}, and a landscape with $K$ local optima must contain at least $K-1$ RSE motifs \citep{riehl2022occurrences,saona2022relation}. This establishes RSE as a necessary local signature of ruggedness.

However, these results do not provide a quantitative prediction for the number of peaks. A landscape may contain many RSE motifs organised in a way that generates a large number of peaks or in a way that gives rise to fewer peaks. Conversely, SSE can be common and strongly affect evolutionary trajectories without necessarily creating additional peaks \citep{Ribeca2026}. Thus, a full theory of peak number must account not only for the amount of RSE but also for how epistatic interactions are arranged.

Another argument comes from the distribution of beneficial mutations. Peaks are genotypes with zero beneficial outgoing mutations, and sinks are genotypes from which all mutations are beneficial. The mean number of beneficial mutations across genotypes is fixed by symmetry: averaged across the whole landscape, half of all mutational directions are beneficial and half are deleterious. By contrast, the variance of this distribution is controlled by local epistasis and is linked to the amount of RSE (see the appendix in \citep{ferretti2018evolutionary} and Ferretti et al, in preparation). As the distribution broadens, more probability mass can appear at the extremes, including genotypes with no beneficial outgoing mutations (peaks) and genotypes from which all mutations are beneficial (sinks). This provides an intuitive connection between local epistasis and peak number, but it still does not yield a general formula.

A direct result comes from the expected number of optima in landscape models such as Rough Mount Fuji \citep{neidhart2014adaptation} or NK \citep{hwang2018universality}, which is directly related to epistatic measures. However, due to the one-parameter structure of these models, these measures do not separate local epistasis from roughness. We will now show that more convincing evidence can be found in the broad family of unstructured Gaussian landscapes, where the expected number of peaks can be computed directly as a function of $\gamma$ or other expected local epistatic measures such as $\gamma^*$ and $\phi_{rs}$, and does not depend on the other $L-2$ parameters describing epistatic interactions in these models.

\subsection{The number of local optima for unstructured landscapes}

In this section, we show how the number of local peaks in unstructured landscape models depends on a single measure of local epistasis, namely the correlation of fitness effects $\gamma$. 

We denote the standard normal distribution by $\varphi(x)$ and its cumulative by $\Phi(x)$. For biallelic landscapes of size $L$ and epistasis parameter $\gamma$, the expected number of local optima is:
\begin{equation}
\mathbb{E}[N_{\mathrm{peaks}}] = 2^L\int_{-\infty}^\infty dx\varphi(x)
\Phi\left(x\sqrt{\frac{1-\gamma}{1+\gamma}}\right)^L.
\end{equation}
irrespective of the details of the power spectrum.
Note that this equation had already been obtained for Gaussian-distributed mean-field NK models in \cite{hwang2018universality}, since they also belong to the family of unstructured Gaussian models. In Supplementary, we show how this equation is generally valid for any unstructured Gaussian model.

\begin{figure}
\centering
\includegraphics[width=0.5\textwidth]{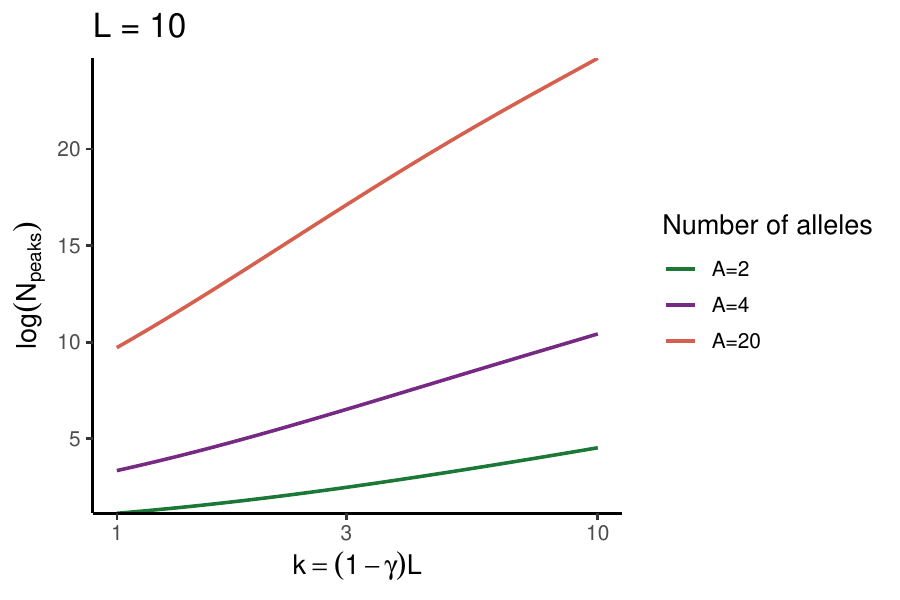}
\caption{Expected number of local optima for unstructured landscapes with $L=10$ as a function of the effective number of epistatic interactions per locus $k=(1-\gamma)L$.}
\label{fig_unstructured_L10}
\end{figure}

\begin{figure}
\centering
\includegraphics[width=0.5\textwidth]{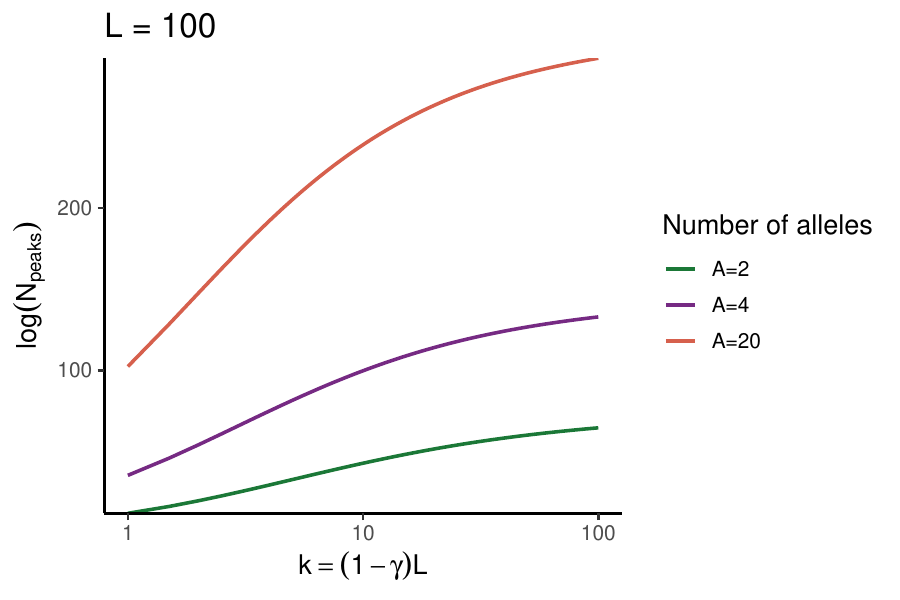}
\caption{Expected number of local optima for unstructured landscapes with $L=100$ as a function of the effective number of epistatic interactions per locus $k=(1-\gamma)L$.}
\label{fig_unstructured_L100}
\end{figure}

It can also be equivalently expressed in terms of other statistics, such as the fraction of reciprocal sign epistasis, using the relation (\ref{eq:rs}):
\begin{equation}
\mathbb{E}[N_{\mathrm{peaks}}] = 2^L \int_{-\infty}^\infty dx\varphi(x)
\Phi\left(x\sqrt{\frac{\sin(\pi\phi_{rs}/2)}{1-\sin(\pi\phi_{rs}/2)}}\right)^L.
\end{equation}
The behaviour of the number of optima as a function of $\gamma$ (or $\phi_{rs}$, since they are approximately proportional) is illustrated in Figures~\ref{fig_unstructured_L10} and \ref{fig_unstructured_L100}.

This expression demonstrates that, in unstructured models with fixed $L$, the expected number of local optima always increases monotonically with $1-\gamma$ (or $\phi_{rs}$). Thus, stronger local epistasis produces more rugged landscapes. However, for fixed $\gamma$, the density of peaks always decreases with the length of the sequence $L$, even though the absolute number of peaks may increase.

It is also illustrative to derive the large $L$ behaviour of this model. We consider the large-$k$ regime in which the effective number of strong interactions per site, $k=L(1-\gamma)$, remains fixed as $L\rightarrow\infty$, and then examine the asymptotic behaviour for $k\gg1$. The density of peaks behaves as 
\begin{equation}
\pi_{\text{peaks}}=\frac{N_{\text{peaks}}}{A^L}\sim e^{\lambda L}\ \ \ \textrm{with}\ \ \ \lambda=-2\frac{\log(k)}{k}
\end{equation}
showing how in this regime, the density of peaks increases with the effective number of interacting sites $k$. The same relation was previously obtained in \cite{hwang2018universality} for mean-field NK landscapes.

For multiallelic landscapes, the corresponding expression depends also on the number of alleles $A$. The multiallelic formula is:
\begin{align}
\frac{\mathbb{E}[N_{\mathrm{peaks}}]}{A^L} & =  \int_{-\infty}^{\infty} dy\varphi(y)
\left[\int_{-\infty}^{\infty} dz\varphi(z)
\Phi\left(\sqrt{1-\gamma}\,y+\sqrt{\gamma}\,z\right)^{A-1}
\right]^L.
\end{align}

\begin{figure}
\centering
\includegraphics[width=0.5\textwidth]{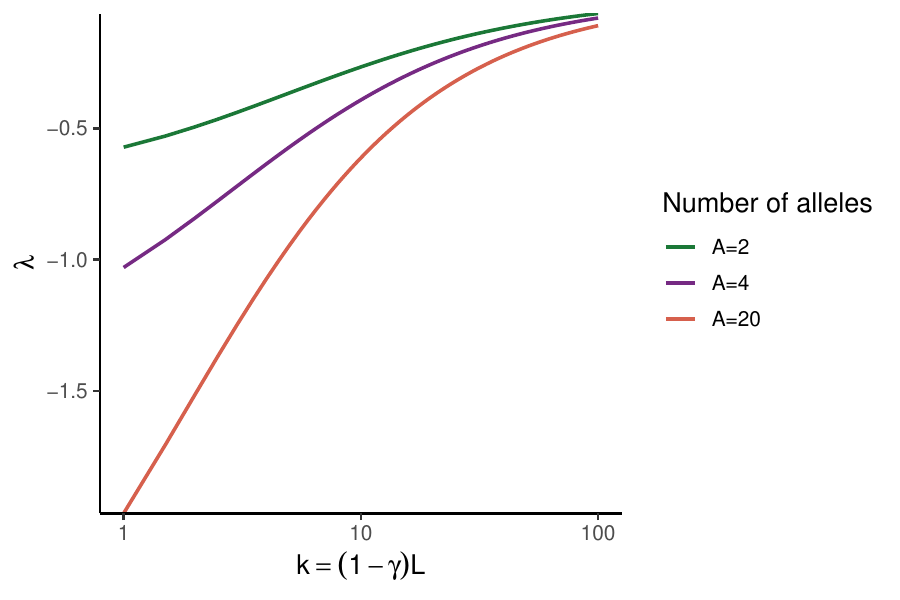}
\caption{Exponential scaling with $L$ of the density of peaks, $\pi_{peaks} \sim e^{\lambda L}$.}
\label{lambda_vs_k_unstructured}
\end{figure}

We use this multiallelic generalisation derived in Supplementary Information to compare peak densities across biallelic, nucleotide-like and amino-acid-like genotype spaces in Figures \ref{fig_unstructured_L10},\ref{fig_unstructured_L100}. In the limit $L\rightarrow\infty$ and $k\gg 1$, this expected value for the density of peaks behaves as 
\begin{equation}
\label{eq:unstruct}
\pi_{\text{peaks}}\sim e^{\lambda L}\ \ \ \textrm{with}\ \ \ \lambda\simeq -2\frac{\log((A-1)k)}{k}
\end{equation}
Exact results for $\lambda$ are shown in Figure \ref{lambda_vs_k_unstructured}. These results imply that for unstructured landscapes, the density of local optima always decreases with the number of alleles $A$. However, the number of local optima may instead increase with $A$ \citep{srivastava2023alphabet}.

\subsection{The number of local optima for real-world landscapes}
We analyse the number of optima for a collection of experimentally characterised biallelic landscapes. Although individual landscapes show substantial sampling noise because they contain only a small number of genotypes and peaks, averages across landscapes with similar local epistasis can still be compared with theoretical expectations. 

\begin{figure}
\centering
\includegraphics[width=\linewidth]{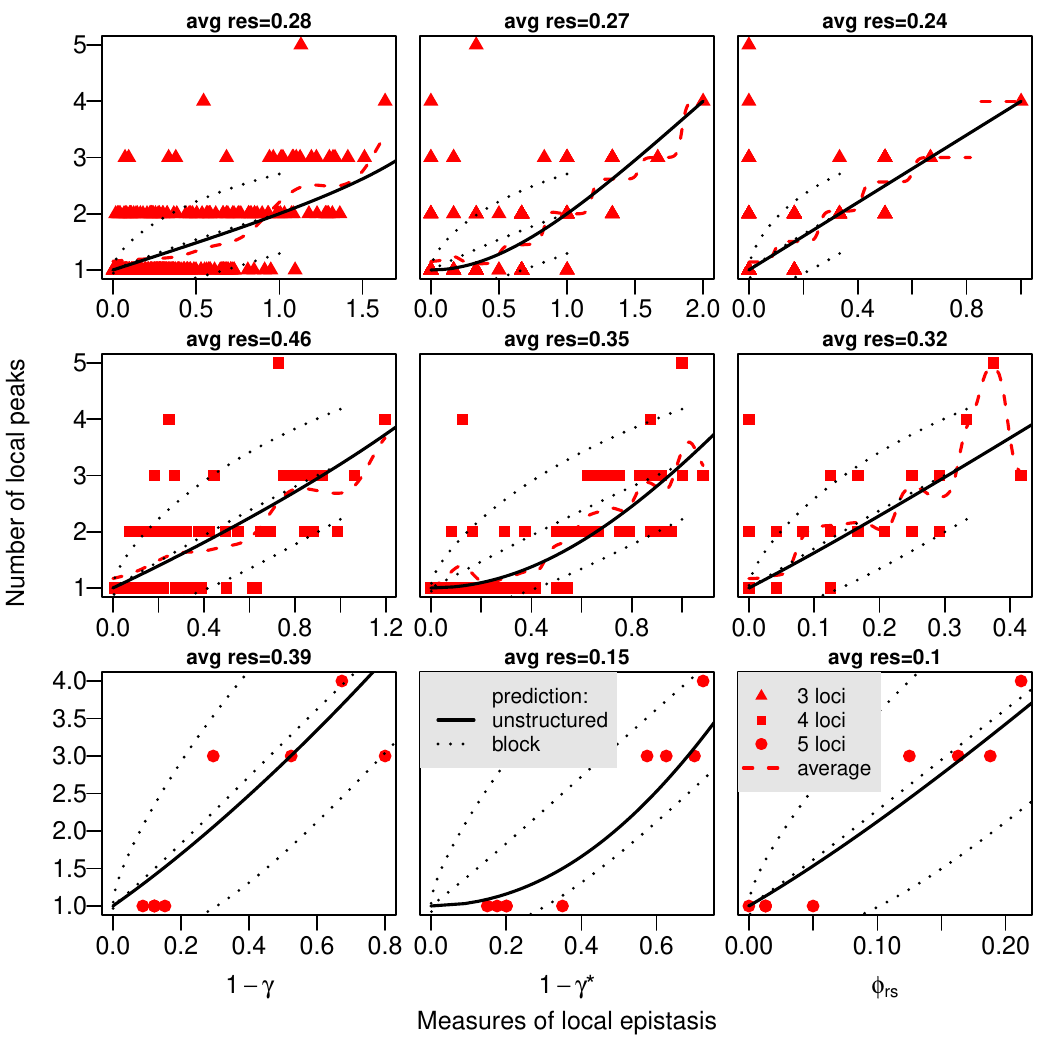}
\caption{Number of maxima versus measures of local epistasis in experimentally resolved (sub)landscapes of different sizes. The dashed lines show a smoothed average of the data, while the continuous lines represent our prediction for unstructured landscapes, with the averaged squared residuals (i.e. mean squared error of the fit) reported above each plot. The dotted lines represent the expected value and 1-sigma interval for the block model.
}
\label{fig_peakemp}
\end{figure}

Interestingly, this collection of empirical landscapes follows quite closely the expected value for unstructured Gaussian models, not only for sign epistasis \citep{Ribeca2026}, but remarkably enough, also for the average number of local peaks, as illustrated in Figure \ref{fig_peakemp}.

Note that the small size of the landscapes ($L=3 - 5$) and large noise associated to individual realisations makes a proper comparison difficult: the predictions for the number of local optima for a very different model, namely the block model \citep{orr2006population,perelson1995protein,schmiegelt2014evolutionary} discussed in the next section as the case of extreme clustering of interactions, would provide a good overall fit for $N_{peaks}$ as well, although unstructured models provide a slightly better fit for some statistics (e.g. for $N_{peaks}$ versus $\gamma^*$ for $L=4,5$).  

\subsection{Heterogeneity and clustering of epistatic interactions impact the number of local optima}

In this section, we examine how heterogeneity between loci and concentration of epistatic interactions affect the number of local peaks, compared to unstructured models.

\subsubsection{Clustering of interactions:} 
For large $L$, clustered interactions with a typical cluster size $1/\beta$ lead to an exponential dependence of the density of peaks on $L$ of the form $\pi_{peaks}\sim e^{\lambda L}$. This depends on the clustering parameter $\beta$ as
\begin{equation}
\lambda=\beta\log\left[\int_{-\infty}^\infty dy\varphi(y)H(y)^{1/\beta}\right]
\end{equation}
where
\begin{align}
     H(y) & = \int_{-\infty}^{\infty} dz \varphi(z)  \Phi\left(y\sqrt{1-\gamma_b}+z\sqrt{\gamma_b}\right)^{A-1} \\
    \gamma_b & = 1-(1-\gamma)\frac{(L-1)\left(\beta+\frac{1}{L-1}\right)}{1-\beta}
\end{align}

\begin{figure}
\centering
\includegraphics[width=0.5\textwidth]{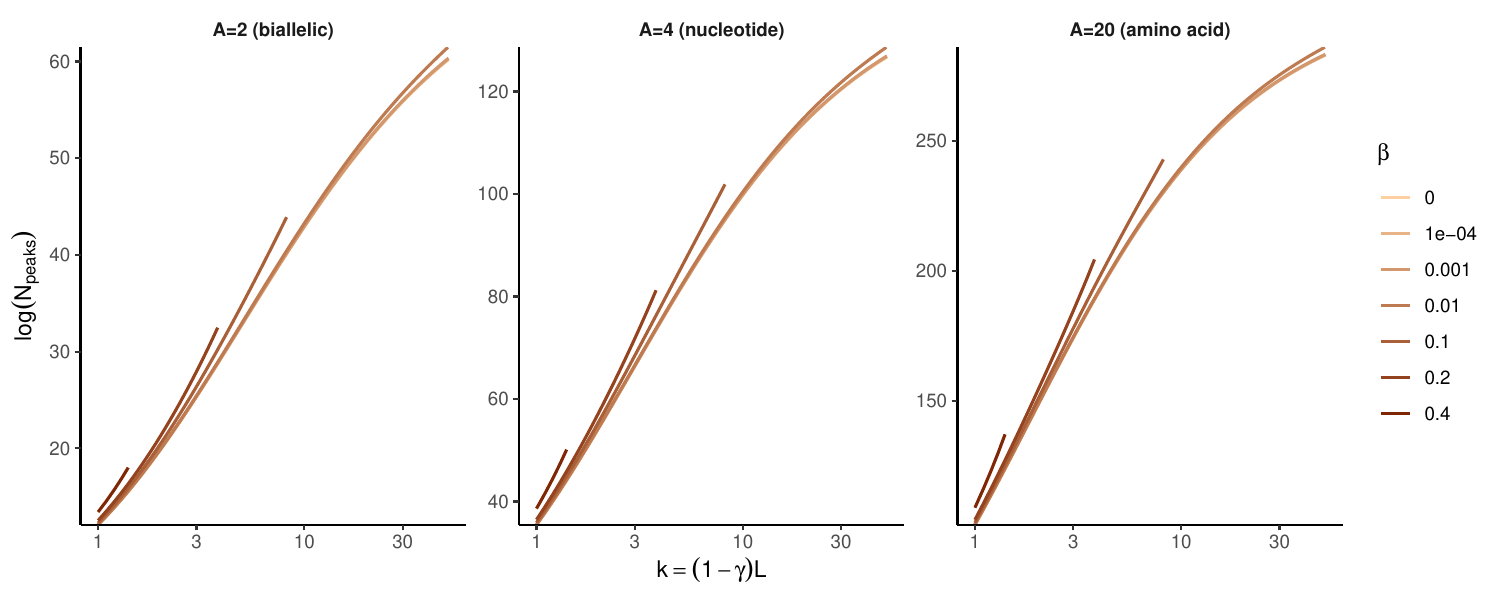}
\caption{Expected number of local optima for landscapes with clustered interactions with $L=100$ as a function of the effective number of epistatic interactions per locus $k=(1-\gamma)L$.}
\label{log_npeaks_clust_L100}
\end{figure}
\begin{figure}
\centering
\includegraphics[width=0.5\textwidth]{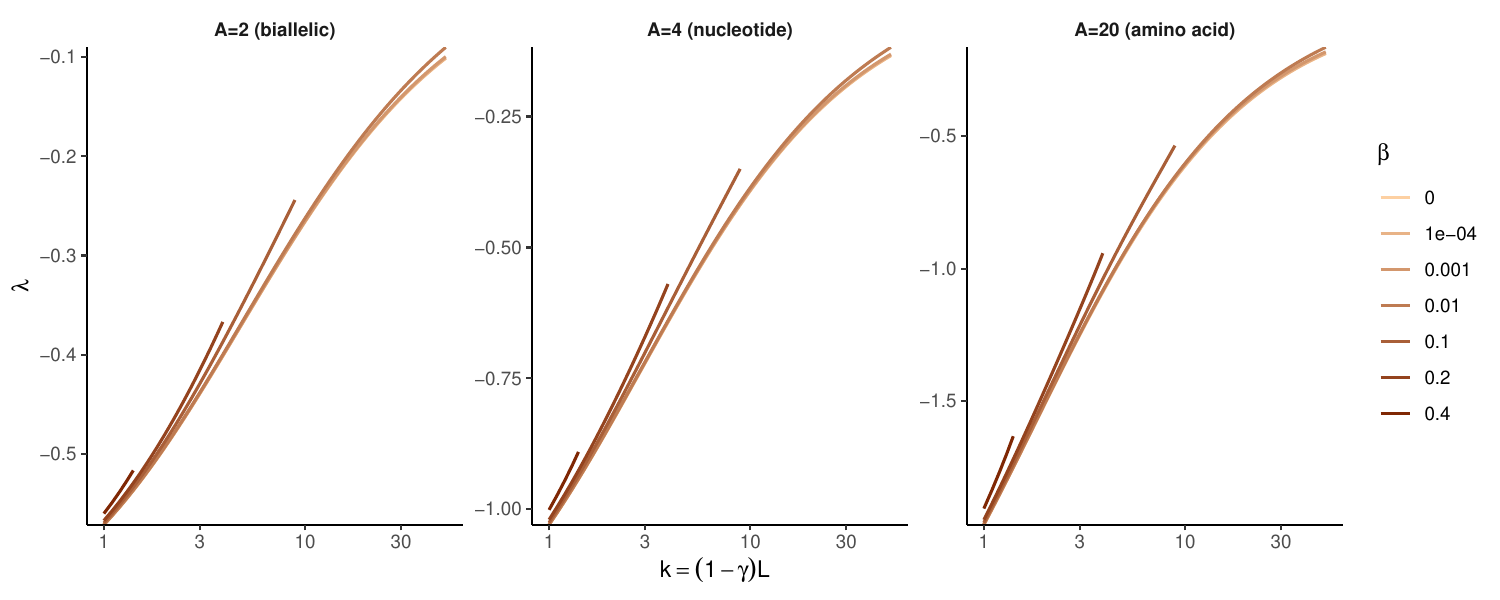}
\caption{Exponential scaling with $L$ of the density of peaks, $\pi_{\text{peaks}} \sim e^{\lambda L}$ for clustered interactions.}
\label{lambda_clust}
\end{figure}

The extreme case of such clustered interactions corresponds to $\beta_{\text{max}}=1/k$, i.e. the limit of a classic block model, which for $k\gg 1$ corresponds to the scaling 
\begin{equation}
\label{eq:max.clust}
\lambda_{\text{max.clust}}\rightarrow\lambda_{\text{block}}=-\frac{\log((A-1)(k+1)+1)}{k+1}.
\end{equation}
This is much larger than the corresponding exponential rate for the unstructured model (by an additional factor of approximately $+\log((A-1)k)/k$). However, as seen in Figures \ref{log_npeaks_clust_L100} and \ref{lambda_clust}, the difference in log-scale is relatively small compared to the potential range in number of optima (which spans the whole interval from $O(1)$ to $O(A^L)$, or from $\lambda\sim -\log(A)$ to $0$). Hence, clustering of interactions matters, but its impact is actually limited.

\subsubsection{Heterogeneity between loci:} 
If there is a fraction $\frac{h}{1+h}$ of loci that do not interact epistatically, the exponential dependence of the density of peaks on $L$, $\pi_{peaks}\sim e^{\lambda L}$ becomes a function of the heterogeneity parameter $h$:
\begin{equation}
\lambda\simeq-\frac{h}{1+h}\log A-2\frac{\log((A-1)k/(1+h)^2)}{(1+h)^3k}
\end{equation}
for $k\gg 1$.

\begin{figure}
\centering
\includegraphics[width=0.5\textwidth]{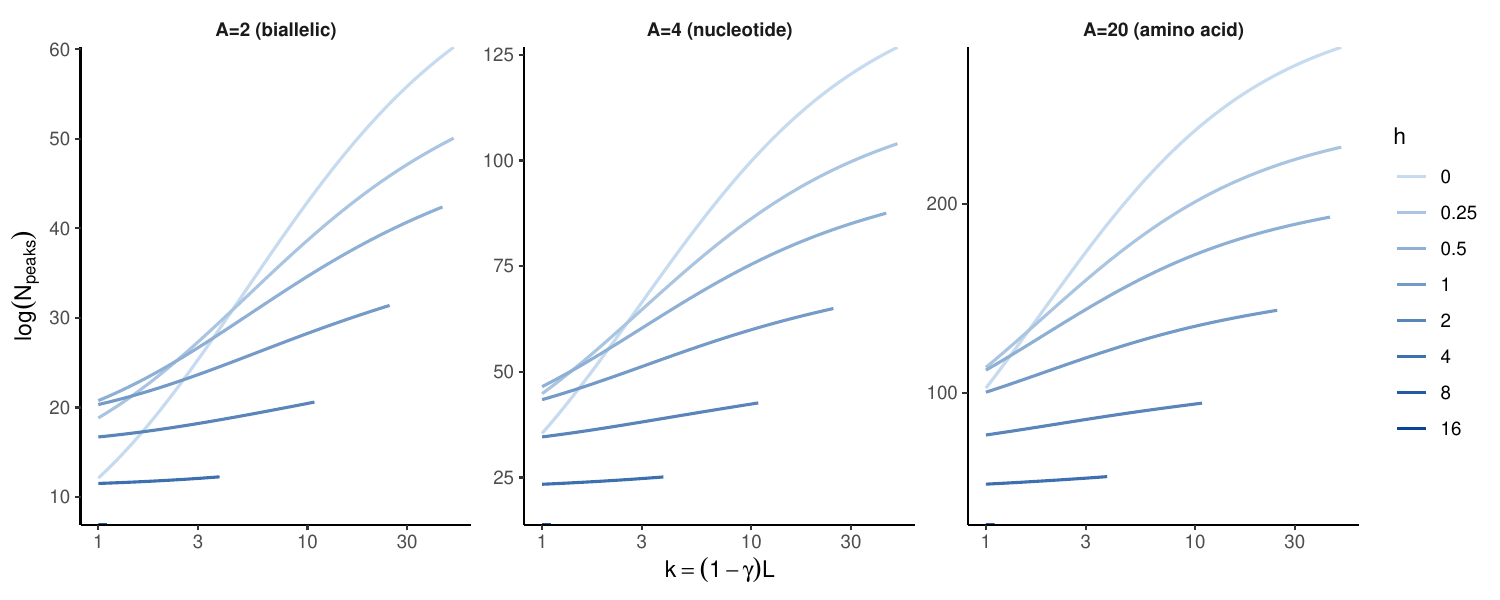}
\caption{Expected number of local optima for landscapes with heterogeneity between loci with $L=100$ as a function of the effective number of epistatic interactions per locus $k=(1-\gamma)L$.}
\label{log_npeaks_het_L100}
\end{figure}

For finite values of heterogeneity $h$, the number of optima can increase or decrease compared to unstructured landscapes (Figure \ref{log_npeaks_het_L100}), due to a combination of two different effects pulling in different directions: the effective reduction in size of the rugged part of the landscape, and the increase in strength of the interactions there. With moderate epistasis and weak heterogeneity, the latter effect is more relevant, resulting in an increase in the number of optima. However, with strong epistasis and/or strong heterogeneity, the former dominates and the number of local peaks is significantly suppressed.

\begin{figure}
\centering
\includegraphics[width=0.5\textwidth]{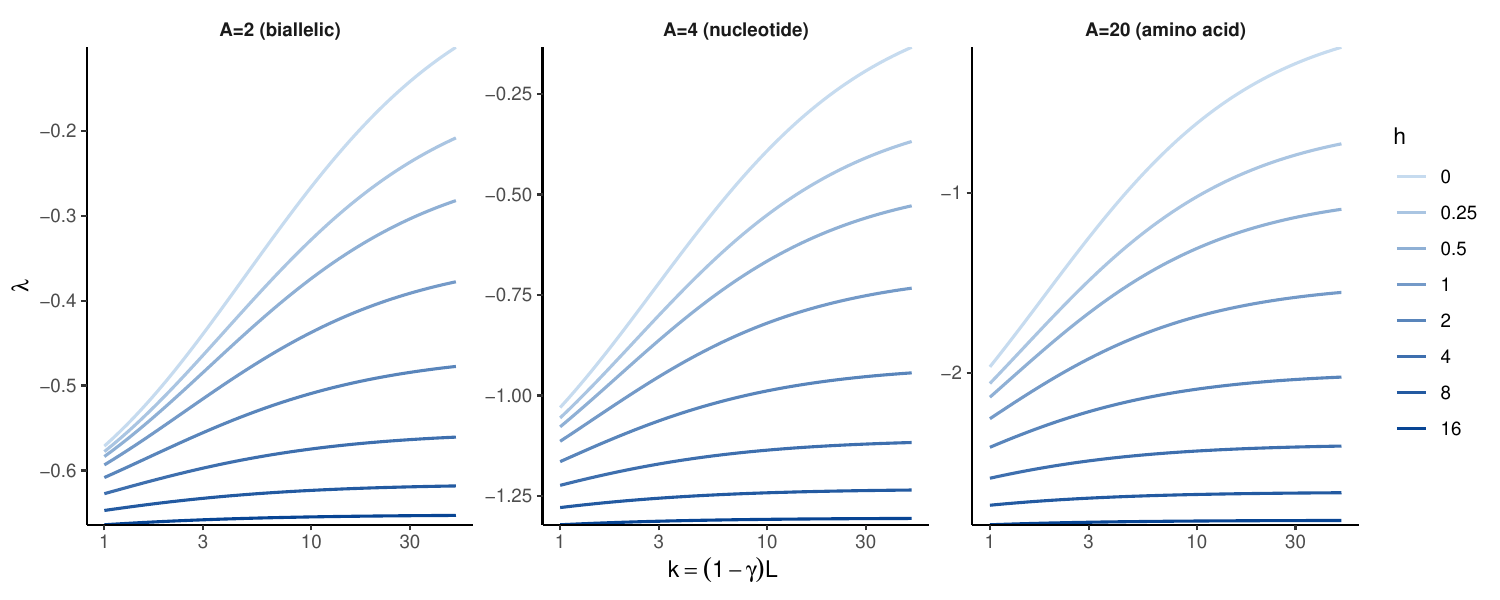}
\caption{Exponential scaling with $L$ of the density of peaks, $\pi_{\text{peaks}} \sim e^{\lambda L}$ for landscapes with heterogeneity between loci.}
\label{lambda_het}
\end{figure}

The most extreme heterogeneity corresponds to $h_{max}=(1-\gamma)^{-1/2}-1\approx \sqrt{L/k}$, i.e. the case where only a fraction $O(1/\sqrt{L})$ of sites participate in the interaction. This extreme scenario actually results in an extreme suppression in the density of peaks (Figure \ref{lambda_het}):
\begin{equation}
\lambda_{\text{max.het}}=-\log(A)
\end{equation}
such that the true scaling of the number of peaks is not exponential anymore in $L$, but in $\sqrt{L}$:
\begin{equation}
\pi_{\text{peaks}}\sim e^{\sqrt{L}\left[\sqrt{k}\log A\right]}.
\end{equation}
There are landscapes with peculiar epistatic structures that are even more extreme in this respect, such as NK models with star structure \citep{hwang2018universality} where the number of peaks remains finite for $L \to \infty$.

\subsubsection{Full model:} 
\begin{figure}
\centering
\includegraphics[width=0.5\textwidth]{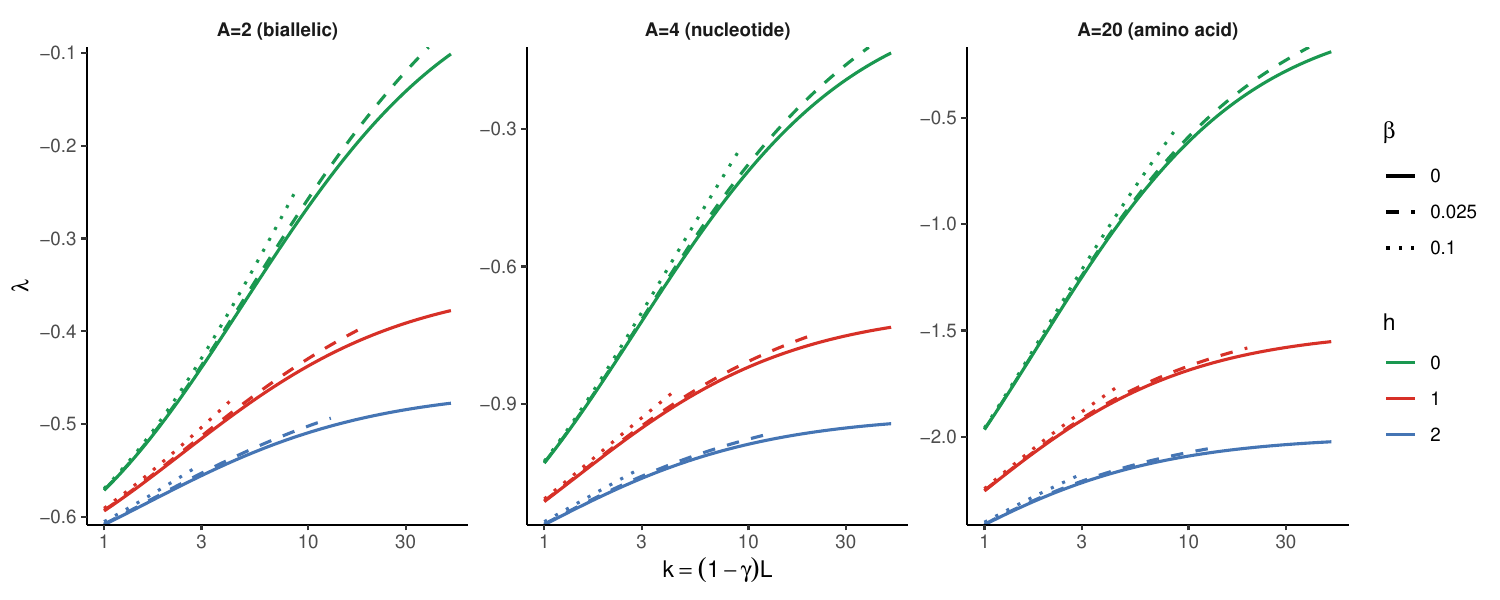}
\caption{Exponential scaling with $L$ of the density of peaks, $\pi_{\text{peaks}} \sim e^{\lambda L}$ for landscapes with both clustering of interactions and heterogeneity between loci.}
\label{lambda_het_clust}
\end{figure}

The formula for the expected number of peaks with both clustering of interactions and heterogeneity between loci is
\begin{align}
    \mathbf{E}[N_{\text{peaks}}] & =A^{\frac{L}{1+h}} 
    \left[ \int_{-\infty}^{\infty} dy \varphi(y) H(y)^{\frac{L}{B(1+h)}}\right]^{B}\label{eq:het_loc_max_beta}
     \end{align}
where 
\begin{align}
     H(y) & = \int_{-\infty}^{\infty} dz \varphi(z)  \Phi\left(y\sqrt{1-\gamma_b}+z\sqrt{\gamma_b}\right)^{A-1} \\
    B & = \beta\left(\frac{L}{1+h}-1\right)+1 \\
    \gamma_b & = 1-(1-\gamma)\frac{(L-1)(1+h)\left(\beta+\frac{1+h}{L-1-h}\right)}{1-\beta}
\end{align}
and the exponential dependence of the density of peaks on $L$ is $\pi_{\text{peaks}}\sim e^{\lambda L}$ with
\begin{equation}
\lambda=-\frac{h}{1+h}\log A+\frac{\beta}{1+h}\log\left[\int_{-\infty}^\infty dy\varphi(y)H(y)^{1/\beta}\right]
\end{equation}
and it is shown in Figure \ref{lambda_het_clust}.

\section{Discussion}

\subsection{What is the typical number of local fitness peaks for typical landscapes?}


The number of local optima is one of the most important topographical features of a fitness landscape. It is an important factor determining to what extent fitness-increasing evolution is constrained by local peaks and therefore shapes the accessibility and predictability of adaptation \citep{srivastava2026}. In this study, we have shown that the typical number of local optima is determined by three properties of epistatic interactions: their strength, their clustering and their heterogeneity across loci.

Our first result is that, in unstructured Gaussian random-field landscapes, the expected number of local optima depends only on the local epistasis parameter $\gamma$. This result provides a simple baseline for predicting ruggedness from the correlation of fitness effects. It also clarifies why local epistasis is more informative than global roughness. Global roughness measures the amplitude of non-additive variation, but local optima depend on the signs and correlations of mutational effects in local neighbourhoods. The same amount of global non-additive variation can therefore produce very different numbers of peaks depending on how it affects local fitness effects.

Our second result is that interaction structure can modulate the number of peaks even when the average strength of epistasis is fixed. Clustering interactions into blocks increases the number of local optima, approaching the block-model limit in which the density of peaks scales as the square root of the unstructured density (at least for many interacting loci), as can be seen by comparing Eqs.~(\ref{eq:unstruct}) and (\ref{eq:max.clust}). This slight increase arises because independent epistatic modules can each generate peaks, and combinations of module-level peaks become peaks of the full landscape.

Our third result is that heterogeneity between loci has an opposite and strong effect. When epistasis is restricted to only a subset of loci, the effective dimensionality of the rugged component of the landscape decreases. Additive loci contribute a single optimum and therefore do not multiply the number of peaks. As a result, heterogeneous landscapes can have far fewer peaks than unstructured landscapes with the same average amount of epistasis.

While we explored a specific class of models, this class is very general and quite flexible. Therefore, we conjecture that the plausible range of peak densities for realistic landscapes is approximately bracketed by:
\begin{itemize}
    \item Upper bound: highly clustered interactions $$\log\pi_{\text{peaks}} \sim -L \frac{\log((A-1)k)}{k}$$
    \item Typical values: unstructured interactions $$\log\pi_{\text{peaks}} \sim -L \frac{2\log((A-1)k)}{k}$$
    \item Lower bound: highly heterogeneous interactions among loci $$\log\pi_{\text{peaks}} \sim -L \log A$$
\end{itemize}
This is consistent with what is known for example from NK models \citep{hwang2018universality}.



\subsection{What is the role of reciprocal sign epistasis?}

These findings refine the relationship between reciprocal sign epistasis and ruggedness. Reciprocal sign epistasis remains the essential local motif required for multiple peaks. In the absence of 
structured epistasis, it is also directly related to the abundance of peaks. In fact, for any unstructured model, we provide a universal equation that links directly the number of local peaks with the fraction of reciprocal sign epistasis. However, the number of peaks is not determined solely by the fraction of reciprocal sign epistatic motifs. Rather, it depends on how those motifs are distributed and connected across the genotype space. This is clear from the role of heterogeneity discussed above, which can dramatically change the density of peaks even when the fraction of reciprocal sign epistasis barely changes. This shows that landscapes with similar amounts of reciprocal sign epistasis can differ substantially in peak number.

\subsection{Conclusion}

The framework developed here has several implications for empirical fitness landscapes. First, the unstructured Gaussian prediction can be used as a null model for the expected number of peaks given the observed amount of local epistasis. This is consistent with the role of these landscapes as natural priors for fitness landscape models \citep{zhou2022higher,zhou2025learning}. Second, deviations from this prediction can be interpreted in terms of interaction structure. Landscapes with more peaks than expected may have clustered or modular interactions, whereas landscapes with fewer peaks may have heterogeneous epistasis concentrated among a subset of loci. Third, because empirical landscapes are usually measured for only a small number of mutations, sublandscape analyses should be interpreted with care: the apparent number of peaks may depend on whether the sampled loci include highly interacting clusters or mostly additive loci.

Several limitations remain. We have focused on Gaussian random-field models and on structured extensions that allow analytical treatment. Real biological landscapes may have non-Gaussian fitness distributions, asymmetric epistatic effects, strong higher-order interactions, multi-allelic constraints or genotype-phenotype maps shaped by biophysical mechanisms and non-linear phenotype-fitness maps. Nevertheless, the distinction between unstructured, clustered and heterogeneous epistasis is general and should apply beyond the specific models studied here. Future work could extend the framework to explicitly mechanistic genotype-phenotype maps, protein and RNA folding landscapes, and machine-learning models of sequence-function relationships.

In summary, the number of local optima in typical fitness landscapes is not determined by epistasis strength alone. For unstructured landscapes, local epistasis provides a universal baseline prediction. Clustering of interactions tends to slightly increase the number of peaks, whereas heterogeneity in interaction participation tends to decrease it. This framework reconciles the necessity of reciprocal sign epistasis with the broader observation that interaction structure determines the ruggedness of a fitness landscape. 

\section{Data availability}
All data analysed in this paper is publicly available.

\section{Acknowledgments}
We thank Sungmin Hwang and Benjamin Schmiegelt for their contributions to the early stages of this work.

\section{Funding}
We acknowledge support from the European REA, Marie Skłodowska-Curie Actions, grant agreement no. 101131463 (SIMBAD). This work was funded by UK Research and Innovation (UKRI) under the UK government’s Horizon Europe funding guarantee (grant number EP/Y037375/1). MG is supported by a Wellcome Trust Early-Career Award (grant 309205/Z/24/Z). JK acknowledges support by Deutsche Forschungsgemeinschaft (DFG) within CRC 1310.

\section{Conflicts of interest}
We declare no conflict of interest.

\bibliography{landscapes}

\end{document}